\documentclass[aps,prl,amssymb,superscriptaddress,10pt,twocolumn,showpacs]{revtex4-1}

\usepackage{amsmath}

\def\d{\mathrm{d}}
\newcommand{\tr}{\operatorname{tr }}
\newcommand{\rmi}{\mathrm{i}}
\newcommand{\rme}{\operatorname{e}}
\DeclareMathOperator{\arcsinh}{arcsinh}

\begin{document}

\title{Positive Hamiltonians can give purely exponential decay}
\date{\today}
\author{Daniel Burgarth}
\affiliation{Department of Mathematics, Aberystwyth University, Aberystwyth SY23 3BZ, UK}
\author{Paolo Facchi}
\affiliation{Dipartimento di Fisica and MECENAS, Universit\`a di Bari, I-70126 Bari, Italy}
\affiliation{INFN, Sezione di Bari, I-70126 Bari, Italy}

\begin{abstract}
It is commonly claimed that only Hamiltonians with a spectrum unbounded both above and below can give purely  exponential decay. Because such Hamiltonians have no ground state, they are considered unphysical. Here we show that Hamiltonians which are bounded below can give  purely exponential decay. This is possible when, instead of looking at the global survival probability, one considers a subsystem only: We show that the reduced state of the subsystem can exhibit a Markovian dynamics and  some (local) observables can decay exponentially. We conclude that purely exponential decay might not be as unphysical as previously thought.
\end{abstract}

\pacs{03.65.-w, 03.65.Xp}


\maketitle

\textit{Introduction.---}Exponential decay is a very general phenomenon across all areas of physics. Examples can be found in high energy physics, in nuclear physics, and in quantum mechanics. The first quantum mechanical theory featuring exponential decay was described by Gamov who considered the alpha decay~\cite{Gamov28}. Other highly influential models were developed by Weisskopf and Wigner for spontaneous emission~\cite{Weisskopf30} and by Friedrichs~\cite{Friedrichs48} and Lee~\cite{Lee56}. 

Despite the success of such models, there is something uncomforting about them: they use Hamiltonians without ground state. Indeed it has been argued that in order to see purely exponential decay, the Hamiltonian \emph{must} be unbounded below and above. Therefore, purely exponential decay is usually considered unphysical. 

For a comprehensive review of the decay process of quantum systems and of the general presence of three regimes---short-time quadratic, exponential, and long-time inverse power law---see~\cite{Nakazato96a}.  See also~\cite{Martorell2009} for an account on long-time behavior.

An influential no-go theorem on Fourier transform  states that the survival probability of the initial state of a quantum system with a finite ground state energy cannot decay exponentially at large times~\cite{Khalfin57,Khalfin58,Exner85}.

It has been part of the folklore that this obstruction holds for a generic observable and also for the reduced dynamics of a subsystem, which has been therefore believed to require an energy spectrum unbounded both above and below in order to be Markovian. This belief has been strengthened also by the fact that in the literature and in textbooks on open quantum systems all known examples  giving an exact  Markovian reduced dynamics at all times involve doubly unbounded Hamiltonians~\cite{Gardiner00,Alicki07}.

Here we will debunk this claim, by explicitly constructing a simple family of exactly solvable positive Hamiltonians on a bipartite system, yielding a purely Markovian dynamics of one of its subsystems, and as a consequence having local observables which decay exponentially.

In doing so we will, in fact, take a natural twist and propose  a change of paradigm in the study of quantum decay of unstable systems: Although the survival probability has been the main object under investigation in this area we suggest to look at other observables too, e.g.\ local observables. 

We will show that a generic observable is sensitive to energy differences, rather than to absolute energy values. Thus the presence of a finite ground state energy becomes ineffective and the observable might decay exponentially. 
In this broader perspective, one realizes that in fact the survival probability---and, in general, a transition probability---is a very peculiar observable decaying according to the absolute spectrum of the Hamiltonian, and one concludes that exponential decay might occur more frequently than expected.

The article is organized as follows. We will first recall the arguments that lead to the conclusion that a quantum system with a finite ground state energy cannot decay exponentially at all times, and then provide a different perspective on the problem by giving a very simple model with positive spectrum, which nevertheless exhibits purely exponential decay.
The eager reader may skip directly to the main result in  the last section on subsystem evolution, which is self-contained.

\vspace{.18cm}

\textit{Quantum behavior at large times.---}Consider a quantum system prepared at $t=0$ in the state $\psi$ belonging to a (complex and separable) Hilbert space $\mathcal{H}$.
Its time evolution is governed by the unitary group $U(t)=\exp(-\rmi t H)$, where $H$ is the Hamiltonian, a (possibly unbounded) self-adjoint operator on $\mathcal{H}$. 

The survival (or nondecay) probability at time $t$ is the square modulus of the survival amplitude~\cite{Nakazato96a}
\begin{equation}
\label{eq:surva}
{a}(t)=\langle \psi | \!\rme^{-\rmi t H}\! \psi \rangle.
\end{equation} 
In terms of the spectral resolution of the Hamiltonian, $H= \int E\, \d P(E)$,  one gets
\begin{equation}
\langle \psi |\! \rme^{-\rmi t H} \!\psi \rangle = \int \rme^{-\rmi E t} \d\mu_\psi (E), 
\end{equation}
where $\mu_\psi (\Omega) = \langle \psi| P(\Omega) \psi \rangle$ is the probability that the energy of the system be in the (measurable) set $\Omega\subset\mathbb{R}$.

Let us  assume that the Hamiltonian $H$ has only an absolutely continuous
spectrum, formally written in Dirac's notation as $H= \int E |E\rangle \langle E| \d E$.
Then the spectral measure $\mu_\psi$ is  absolutely continuous, namely its derivative 
\begin{equation}
 p_\psi(E) = \frac{\d\mu_\psi(E)}{\d E}
\label{eq:abscont}
\end{equation} 
is an integrable and normalized function of energy~$E$. This derivative
represents the ``energy density'' of state~$\psi$
\begin{equation}
p_\psi(E)=|\langle E | \psi\rangle|^2\geq 0 .
\label{eq:specdens}
\end{equation} 
Therefore, the survival amplitude~\eqref{eq:surva} is
the Fourier transform of the energy density~\eqref{eq:specdens}, as first pointed out by Fock and Krilov~\cite{Fock47}. 

Since $p_\psi(E)$ is an
 integrable function, by  the Riemann-Lebesgue
lemma~\cite{Strichartz94}, the survival amplitude ${a}(t)$ is continuous and vanishes at infinity:
\begin{equation}
\lim_{t\to\infty} {a}(t) = \lim_{t\to\infty}  \int \rme^{-\rmi Et} p_\psi(E) \d E 
=0,
\label{eq:RiemannLebesgue}
\end{equation}
so that $\psi$ is a truly unstable
state.

Assume now on physical grounds that the energy
spectrum is  bounded from below in order to have a finite ground energy.
It follows that the spectral density~\eqref{eq:specdens} vanishes for $E<E_g$, where
$E_g>-\infty$ is the ground energy, that for convenience will be assumed to be $0$, and we get
\begin{equation}
\label{eq:aUaFT1}
{a} (t) = \int_{0}^\infty \rme^{-\rmi Et} p_\psi(E) \d E .
\end{equation}

A theorem of Fourier analysis due to Paley and Wiener~\cite{Paley34} states that
 if $p_\psi(E)$ is an integrable function supported on $[0,+\infty)$, then its Fourier trasform ${a}(t)$ must
satisfy the inequality
\begin{equation}
  \int_{-\infty}^\infty
  \frac{-\ln\vert{a}(t)\vert}{1+t^2} \d t<\infty.
\label{eq:PaleyWiener}
\end{equation}
Therefore, the survival probability $|a(t)|^2$ cannot be an exponential, for
the integral (\ref{eq:PaleyWiener}) would diverge as $\ln t$ as
$t\rightarrow\infty$: At large times the decay must be slower, for
example a power law.

Notice that this is a very general result: the only condition
required is the existence of a finite ground state energy $E_g$. The use of a
Paley-Wiener theorem in this context is due to Khalfin~\cite{Khalfin57,Khalfin58}. A full account of this theorem, and  further references are given in~\cite{Exner85}.

\vspace{.18cm}

\textit{Purely exponential decay.---}Now we seek a Hamiltonian~$H$ and an initial state $\psi$ such that the survival amplitude~\eqref{eq:surva} is exponential for all times, namely
\begin{equation}
\label{eq:exp}
{a}(t)
= \int \rme^{-\rmi E t} \d\mu_\psi (E)
= \exp\left(-\frac{\gamma}{2} |t| - \rmi\, \omega_0 t\right),
\end{equation} 
with $\gamma >0$ and $\omega_0\in\mathbb{R}$.
From the previous discussion we know that $H$ should be unbounded both below and above.

Since the probability measure $\mu_\psi$ is the inverse Fourier transform of an integrable function $a(t)$, 
then it is absolutely continuous, that is of the form~\eqref{eq:abscont}.
Explicitly we get
\begin{eqnarray}
p_\psi(E) = \frac{1}{2\pi} \int_\mathbb{R} \rme^{\rmi E t } {a}(t)\, \d t 
= \frac{\gamma}{2\pi} \frac{1}{(E-\omega_0)^2 + \frac{\gamma^2}{4}},
\label{eq:Lorentzian}
\end{eqnarray}
and $\psi$ is in the absolutely continuous spectral subspace of $H$. 

The (minimal) solution to our problem~\eqref{eq:exp} is therefore a Hamiltonian $H$ with absolutely continuous spectrum on the full line
and an initial state $\psi$, whose wave function  in the energy representation $L^2(\mathbb{R}_E)$ is a square root of the Cauchy-Lorentz function~\eqref{eq:Lorentzian}:
\begin{equation}
\label{eq:phi_a}
 \langle E | \psi\rangle = \phi_{\mathrm{C}}(E) = \sqrt{p_\psi(E)} \, \rme^{\rmi \alpha(E)},
\end{equation}
with $\alpha(E)$ an arbitrary real function.
For example,
\begin{equation}
\phi_{\mathrm{C}}(E) = \sqrt{\frac{\gamma}{2\pi}} \frac{1}{E-\omega_0 - i \frac{\gamma}{2}}.
\label{eq:phi_a2}
\end{equation}

As canonical physical realizations, one can think of the position operator on the line $H=q$ with initial wave function $\langle x | \psi\rangle=\phi_{\mathrm{C}}(x)$,  and one has
\begin{equation}
\label{eq:qevol}
\langle \phi_{\mathrm{C}}| \!\rme^{-\rmi t q}\! \phi_{\mathrm{C}}\rangle = \exp\left(-\frac{\gamma}{2} |t| - \rmi\, \omega_0 t\right).
\end{equation}
Alternatively, one can consider  the momentum operator $H=p$ with $\langle p | \psi\rangle=\phi_{\mathrm{C}}(p)$. Although both operators $q$ and $p$ correspond to physical observables, as Hamiltonians they are considered unphysical since have no stable ground states.

\vspace{.18cm}

\textit{Subsystem evolution.---}Now we will exhibit a simple model of a bipartite quantum system with positive Hamiltonian, which nevertheless has a subsystem with a purely Markovian dynamics and local observables which decay exponentially.
Consider a spin coupled to an unbound system, $\mathcal{H}= \mathbb{C}^2 \otimes L^2(\mathbb{R})$. 
As Hamiltonian consider the following operator
\begin{equation}
\label{eq:H1}
H= |0\rangle \langle 0| \otimes \, q_{+} + |1\rangle\langle 1| \otimes \, q_{-}, 
\end{equation}
where $\{|0\rangle, |1\rangle\}$ is the canonical basis of the spin space~$\mathbb{C}^2$, 
$q_+$ is the position operator on the positive real line, that is the multiplication operator by the ramp function $x_+ =\max\{x,0\} = x\, \theta (x)$:
\begin{equation}
q_+ \phi(x) = x_+ \phi(x),
\end{equation}
and $q_-$ is the multiplication operator by the ramp function $x_- = \max\{-x,0\} = -x \, \theta (-x)$.
This is a variation of the ``shallow pocket model''~\cite{Arenz14}.

The operators $q_{\pm}$ are self-adjoint on their maximal domains, 
$D(q_{\pm})= \{ \phi \in L^2(\mathbb{R}), q_{\pm} \phi \in L^2(\mathbb{R})\}$, 
and thus the Hamiltonian $H$ is self-adjoint on $D(H) =  D(q_+) \oplus D(q_-)$. 
Moreover, it is straightforward to prove that
\begin{equation}
q = q_+ - q_-,
\label{eq:diff}
\end{equation}
where $q$ is the position operator on its natural domain.

The crucial property we want to  emphasize is that  $H$ is positive, that is
\begin{eqnarray}
& & \langle \psi | H \psi\rangle = \langle \psi_1 | q_+ \psi_1 \rangle + \langle \psi_2 | q_- \psi_2 \rangle
\nonumber\\
& &  = \int_0^{+\infty} x (|\psi_1(x)|^2 + |\psi_2(-x)|^2) \d x 
 \geq 0 , 
\end{eqnarray}
for all $\psi = (\psi_1,\psi_2)\in D(H)$. 

The unitary group $U(t)=\rme^{-\rmi t H}$ generated by the Hamiltonian~\eqref{eq:H1} is
\begin{equation}
U(t) = |0\rangle \langle 0| \otimes  \rme^{-\rmi t q_{+}} + |1\rangle\langle 1| \otimes  \rme^{-\rmi t q_{-}} .
\label{eq:U(t)}
\end{equation}
Now take an initial factorized state $\rho \otimes |\phi_{\mathrm{C}}\rangle\langle \phi_{\mathrm{C}}|$, with $\rho$ an arbitrary spin density matrix and $\phi_{\mathrm{C}}$ given by~\eqref{eq:phi_a2}, and look at the evolution of the spin reduced state, given by a partial trace over the unbound subsystem:
\begin{equation}\label{subsystem}
\rho(t) = \tr_2 \left(U(t) (\rho \otimes |\phi_{\mathrm{C}}\rangle\langle \phi_{\mathrm{C}}| ) U(t)^\dagger \right).
\end{equation}
By using~\eqref{eq:diff} and~\eqref{eq:qevol}, one easily gets
\begin{eqnarray}
\rho(t) &=& \rho_{00} |0\rangle \langle 0| + \rho_{11} |1\rangle \langle 1|
\nonumber\\
& & +  f(t) \rho_{01} |0\rangle \langle 1| + \overline{f(t)}  \rho_{10} |1\rangle \langle 0| ,
\label{eq:ampldamp}
\end{eqnarray}
where $\rho_{ij} =  \langle i| \rho |j\rangle$ and 
\begin{equation}
f(t) = \langle \phi_{\mathrm{C}}| \! \rme^{-\rmi t q}\! \phi_{\mathrm{C}}\rangle = \rme^{-\frac{\gamma}{2} |t| -\rmi\, \omega_0 t}.
\label{eq:fexp}
\end{equation}
This is nothing but a phase damping channel on the spin. 
The reduced dynamics is Markovian
\begin{equation}
\label{eq:Markovian}
\rho(t) = \rme^{t \mathcal{L}}\! \rho,
\end{equation}
with GKLS generator~\cite{GKS76,Lind76}
\begin{equation}
\mathcal{L}\rho(t)=-\rmi\,\omega_0 [\sigma_z,\rho(t)]-\frac{\gamma}{2}[\sigma_z,[\sigma_z,\rho(t)]],
\label{eq:GKLS}
\end{equation} 
where $\sigma_z=|0\rangle\langle 0|-|1\rangle\langle 1|$.

Therefore, although the Hamiltonian~(\ref{eq:H1}) is positive, it generates a purely Markovian dynamics on the spin subsystem!
And in particular, if one looks at the time evolution of the polarization $\sigma_x = |0\rangle\langle 1|+|1\rangle\langle 0|$,
one gets that, for $\omega_0=0$, it decays exponentially:
\begin{equation}
\langle \sigma_x(t) \rangle = \tr (\sigma_x \rho(t) ) = 
\langle \sigma_x(0)\rangle  \rme^{-\frac{\gamma}{2} |t|}.
\end{equation}

This does not contradict the Fourier analysis argument given in the previous sections. Essentially, when considering a subsystem evolution~(\ref{subsystem}) one needs to take the adjoint action of the unitary on the density matrix of the total system, $U(t)\rho_{\mathrm{tot}} U(t)^\dagger$. The corresponding generator (superoperator)~$\mathcal{G}$ acts as~$\mathcal{G} \rho_{\mathrm{tot}} = [H,\rho_{\mathrm{tot}} ]$, and has as a spectrum the \emph{differences} of the spectrum of $H$. Therefore, it is unbounded  above \emph{and} below, and thus is potentially capable of producing an exponential decay.

This argument applies to a generic observable $A=A^\dagger$, whose expectation value evolves as
\begin{equation}
\label{eq:A(t)}
\langle A (t) \rangle = \tr ( A \!\rme^{-\rmi t H}\! \rho_{\mathrm{tot}}  \! \rme^{+\rmi  t H} )= \tr\!\left( A \rme^{-\rmi  t \mathcal{G}} (\rho_{\mathrm{tot}})\right).
\end{equation}
However, it becomes ineffective for the survival probability (or in general for a transition probability) of a global pure state~$\psi$. Indeed, in such a case 
$A=\rho_{\mathrm{tot}} = |\psi\rangle \langle\psi|$, and~\eqref{eq:A(t)} reduces to the square modulus of the survival amplitude $a(t)$ which by~\eqref{eq:surva} depends only on the Hamiltonian~$H$ and its spectrum: The above argument is defused and the lower boundedness of the Hamiltonian and the Paley-Wiener theorem go into action.

In fact, with our Hamiltonian~\eqref{eq:H1} more can be said. It is easy to see from~\eqref{eq:U(t)} that for a pure initial state $\psi = \chi \otimes \phi_{\mathrm{C}}$ one gets 
\begin{equation}
a(t) = \langle \psi | U(t) \psi \rangle  \to (|\langle 0 | \chi\rangle|^2 +  |\langle 1 | \chi\rangle|^2)/2,
\end{equation}
as $t\to \infty$, since $\langle \phi_{\mathrm{C}}|\!\rme^{-\rmi t q_{\pm}}\! \phi_{\mathrm{C}}\rangle \to 1/2$ (with a non-exponential behavior). Thus, the Markovian dynamics~\eqref{eq:Markovian} and the exponential decay of $\sigma_x$ take place, notwithstanding the global state $\psi$ is  not even truly unstable, and its survival probability  does not fully decay.

Finally, notice that non-analiticity at 0 of the ramp function~$x_+$ is inessential.
Indeed, in~\eqref{eq:H1} one can replace the positive operators $q_{\pm}$ with the multiplication operators by the functions $V(\pm x)$:
\begin{equation}
\label{eq:H2}
H= |0\rangle \langle 0| \otimes \, V(q) + |1\rangle\langle 1| \otimes \, V(-q),
\end{equation}
where $V(x)$ is an arbitrary absolutely continuous non-negative monotone function which is strictly monotone on the positive real line.
For example, one can take $V(x)= \exp(x)$, which is entire analytic.

One gets that $W(x)= V(x) - V(-x)$ is a strictly monotone function with domain and range $\mathbb{R}$, and thus is invertible. The reduced time evolution of a factorized initial state 
$\rho \otimes |\phi\rangle\langle \phi|$ is  given again by~\eqref{eq:ampldamp} with 
\begin{equation}
f(t) = \langle \phi| \rme^{-\rmi t W(q)} \phi \rangle.
\end{equation}
By choosing as initial state
\begin{equation}
\phi(x) = |W'(W^{-1}(x))|^{-1/2} \phi_{\mathrm{C}}(W^{-1}(x)),
\end{equation}
which is a normalized square-integrable function, then $f(t)$ is easily proved to be exponential as in~\eqref{eq:fexp}.

For example, when $V(x)= \exp(x)$, then $W(x)= 2 \sinh x$ and the initial wave function giving exponential decay reads
\begin{equation}
\phi(x) = \sqrt{\frac{\gamma}{4\pi  \cosh y}} \, \frac{1}{y -\omega_0 - i \frac{\gamma}{2}},
\end{equation}
with $y=\arcsinh(x/2)$.

\vspace{.2cm}

\textit{Conclusions.---}We have given a very simple model of purely exponential decay caused by a positive Hamiltonian. It shows that the decay properties of a subsystem, i.e.\ of a factor of a tensor product, can be very different from those 
of the total system or of a component of a direct sum,  which has so far been the  framework for the study of unstable systems. 

In the theory of open quantum systems, on the one hand usually a Markovian master equation for a subsystem is obtained by using delicate approximations or limiting procedures~\cite{Breuer07}. On the other hand, 
in the models proposed in the literature, reservoirs without such approximations but with purely exponential decay are extreme idealizations of real systems with  singular couplings and 
unbounded from below Hamiltonians~\cite{Gardiner00,Alicki07, Waldenfels14}. 
We have shown that the latter idealization, at least in some cases, is in fact unnecessary,
and thus Markovianity and purely exponential decay might not be as unphysical as previously thought.

The example should stimulate further mathematical and physical studies of purely exponential decay by positive Hamiltonians, and provide a substantial contribution to the discussion of exponential decay in quantum physics. For the sake of argument we conclude with some open problems which naturally arise, among others, from our findings and that are worth investigating in the future.
Given a positive Hamiltonian, which observables exhibit purely exponential decay? Must they be local or do there exist genuine global observables decaying exponentially? Which type of Markovian reduced dynamics one can get?  Notice that the  dynamics obtained in~\eqref{eq:GKLS} is a phase damping channel. Can the exponential decay out of a bounded below Hamiltonian arise with different types of channels as well, or there is something special about  dephasing?

\section{acknowledgements}

We acknowledge discussions with  Marilena Ligab\`o.
DB acknowledges support by the EPSRC Grant No. EP/M01634X/1.
PF was partially supported by INFN through the project ``QUANTUM'', and by the Italian National Group of Mathematical Physics (GNFM-INdAM).

\bibliographystyle{plain}

\end{document}